\documentclass[a4paper,usenatbib]{mn2e}

\usepackage{amssymb,amsmath}
\usepackage{graphicx}
\usepackage{float}
\usepackage{subfig}
\usepackage{multirow}
\usepackage{aas_macros}
\usepackage{fixltx2e}
\usepackage[usenames]{color}

\title[Accurate positions and optical counterparts of ULXs NGC 7319-X4 and NGC 5474-X1]
{Accurate positions for the ULXs NGC 7319-X4 and NGC 5474-X1 and limiting magnitudes for their optical counterparts}
\author[M. Heida et al.]
{M. Heida,$^{1,2}$ P. G. Jonker,$^{1,2,3}$ M. A. P. Torres,$^{1,3}$ S. Mineo,$^{3}$\\
$^1$SRON Netherlands Institute for Space Research, Sorbonnelaan 2, 3584 CA Utrecht, the Netherlands\\
$^2$Department of Astrophysics/IMAPP, Radboud University Nijmegen, P.O. Box 9010, 6500 GL Nijmegen, The Netherlands\\
$^3$Harvard-Smithsonian Center for Astrophysics, 60 Garden Street, Cambridge, MA 02138, USA} 

\begin{document}
\maketitle

\begin{abstract}
In this paper we report accurate \textit{Chandra} positions for two ultraluminous X-ray sources: NGC 7319-X4 at Right Ascension (RA) = 339.02917(2)$^{\circ}$, Declination (Dec) = 33.97476(2)$^{\circ}$ and NGC 5474-X1 at RA = 211.24859(3)$^{\circ}$, Dec = 53.63584(3)$^{\circ}$. We perform bore-sight corrections on the \textit{Chandra} X-ray Satellite observations of these sources to get to these accurate positions of the X-ray sources and match these positions with archival optical data from the Wide Field and Planetary Camera 2 on board the Hubble Space Telescope. We do not find the optical counterparts: the limiting absolute magnitudes of the observations in the WFPC2 standard magnitude system are B = $-7.9$, V = $-8.7$ and I = $-9.3$ for NGC 7319-X4 and U = $-6.4$ for NGC 5474-X1. We report on the X-ray spectral properties and we find evidence for X-ray variability in NGC 5474-X1. Finally, we briefly discuss several options for the nature of these ULXs.
\end{abstract}

\section{Introduction}
Ultraluminous X-ray sources (ULXs) are pointlike, off-nuclear X-ray sources that have a luminosity that is larger than the Eddington luminosity of a 10 M$_{\sun}$ black hole ($L_{\textrm{Edd}}(10 \textrm{ M}_{\sun}) \approx 10^{39}$ erg s$^{-1}$; \citealt{colbert05}).
Several scenarios have been proposed to explain the high luminosities of these sources. The sources with $L \lesssim 10^{40}$ erg s$^{-1}$ can be explained by stellar mass black holes that emit radiation beamed instead of isotropically \citep{king01}. Alternatively, the black hole can be more massive than the typical 10 M$_{\sun}$ if formed by a low-metallicity progenitor, which would lead to a higher Eddington luminosity \citep{belczynski10}. ULXs with luminosities above $10^{41}$ erg s$^{-1}$, sometimes called hyperluminous X-ray sources, may however need black holes with $M \gtrsim 1000 M_{\sun}$ to explain \citep{colbert05}. Such black holes are referred to as intermediate mass black holes (IMBHs).
Another possibility is that some of these sources could be recoiling supermassive black holes (SMBHs). In the standard $\Lambda$CDM cosmology galaxies grow through hierarchical mergers. The central massive black holes of merging galaxies may eventually merge as well. Simulations show that when two black holes coalesce, gravitational waves are emitted that carry away linear momentum in a preferential direction just prior to the merger. As a result the black hole receives a kick in the opposite direction. The size of the kick depends on the magnitude and orientation of the spins of the black holes before the merger and on their mass ratio (\citealt{damour06}, \citealt{ferrarese02}, \citealt{merritt09}). The maximum kick velocity is of the order of 1000 km/s \citep{baker08}, which, with time, would allow one to observe these recoiled SMBHs as bright off-nuclear X-ray sources \citep{jonker10}.
Another possibility is that some of these ULXs are supernovae. SN type IIn are the X-ray-brightest supernovae and therefore the best candidates \citep{immler03}.
Finally, some ULXs could potentially be explained as the remnant BH from the smaller galaxy that underwent a merger event with a larger galaxy.

Finding the optical counterparts of ULXs can help us distinguish between the different scenarios for their nature. If the ULX is a stellar or intermediate mass black hole we expect to see an accretion disc and/or a companion star. A recoiling SMBH would be surrounded by a cluster of stars \citep{merritt09}, whereas a supernova could be distinguished by its decaying light curve. For several ULXs the optical counterpart has been reported (e.g. \citealt{motch11}, \citealt{yang11}, \citealt{wiersema10}). In this Paper, we investigate two ULXs that were observed by the \textit{Chandra} X-ray Observatory and the Hubble Space Telescope (HST). NGC 7319-X4 \citep{liu11} is located in the late-type galaxy NGC 7319, one of the galaxies in Stephan's Quintet. NGC 5474-X1 \citep{swartz11} is the brightest X-ray source in the highly asymmetrical late-type galaxy NGC 5474.

\section{Observations, analysis and results}
\subsection{\textit{Chandra} X-ray observations}

\begin{table*}
\begin{minipage}{110mm}
\caption{The \textit{Chandra} observations of the ULXs in NGC 7319 and NGC 5474: observation ID number, exposure time in kiloseconds, CCD on which the source was detected and the off-axis angle of the source in arcminutes. }
 \begin{tabular}{cccccc}
 \hline
 Galaxy & Obs. ID & Exp. time & src on CCD & off-axis angle & Obs. date \\
 \hline
\multirow{2}{*}{NGC 7319} & 789 & 19.7 ks & ACIS S3 & 2.00' & 2000-07-09\\
 & 7924 & 94.4 ks & ACIS S3 & 3.05' & 2007-08-17\\
\multirow{2}{*}{NGC 5474} & 7086 & 1.8 ks & ACIS S3 & 1.61' & 2006-09-10\\
 & 9546 & 30.2 ks & ACIS S3 & 1.89' & 2007-12-04\\
 \hline
 \end{tabular}
\label{tab2}
\end{minipage}
\end{table*}

We retrieve the observations in Table \ref{tab2} from the \textit{Chandra} archive and
reprocess the events making use of the calibration files using \textsc{caldb} version 4.4.7, employing \textsc{CIAO} version 4.4. We only use datasets 7924 and 9546 for the bore-sight correction of the \textit{Chandra} images, since the datasets with the shorter exposure times do not contain other sources with enough counts for accurate position measurements. 

\subsubsection{NGC 7319}
\begin{figure*}
 \hbox{
\includegraphics[width=0.4\textwidth]{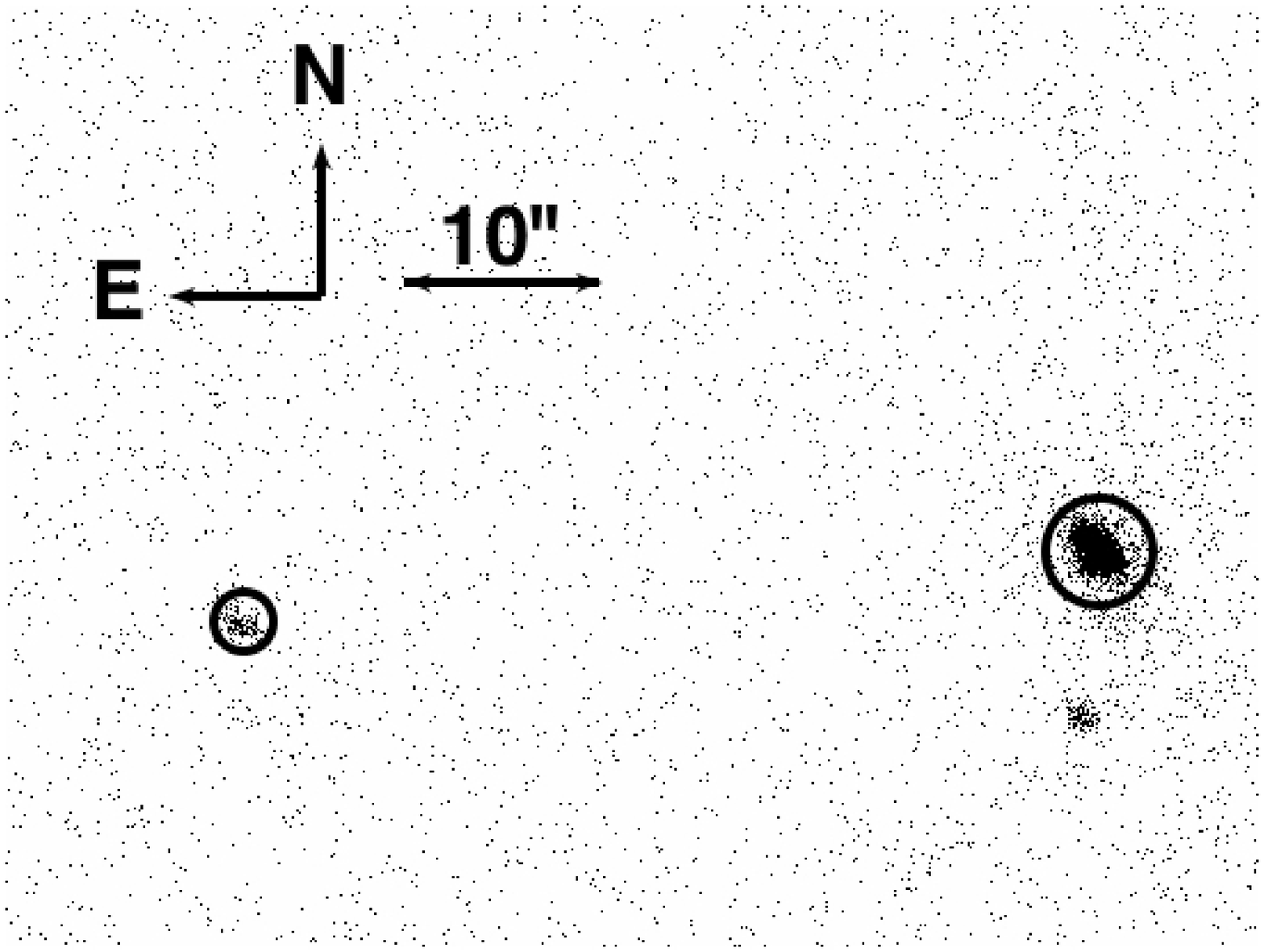}
\includegraphics[width=0.6\textwidth]{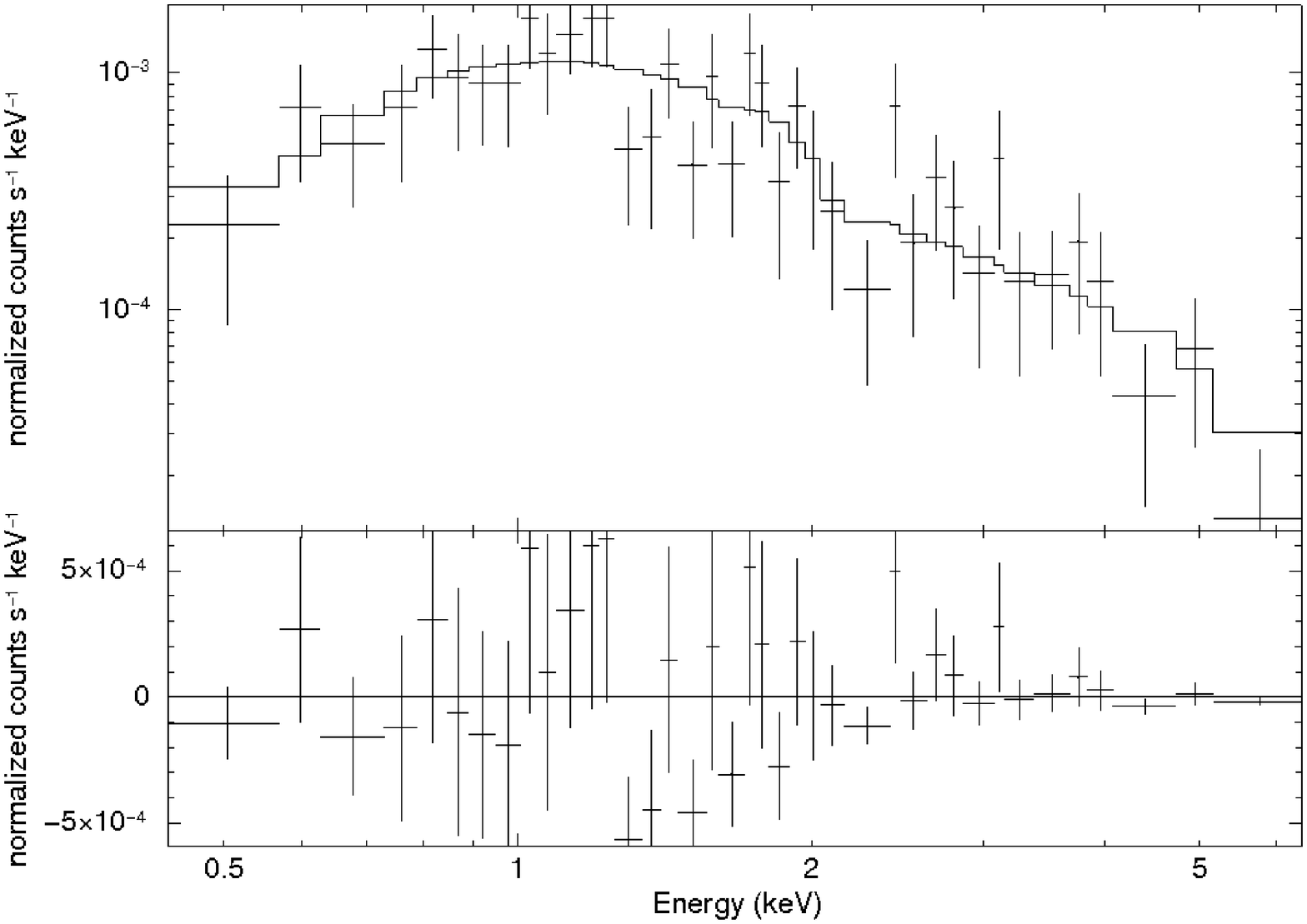}
}
\caption{\emph{Left panel:} The \textit{Chandra} image with the ULX (left circle) and position of the core of NGC 7319 (right circle). \emph{Right panel:} The X-ray spectrum of NGC 7319-X4 with fitted model and residuals. The events are extracted from the observation with ID 7924 (see Table \ref{tab2}) and are rebinned for visual clarity.}\label{chan7319}
\end{figure*}

For the bore-sight correction of the NGC 7319 field we can use only one source, coinciding with the center of the galaxy (see Figure \ref{chan7319}). This is a bright (K-band magnitude = 10) point source in the 2 Micron All Sky Survey (2MASS) and its position is accurately known: Right Ascension (RA) = 339.014833$^{\circ}$ and Declination (Dec) = 33.975750$^{\circ}$.
The position of the X-ray source from \textsc{wavdetect} is RA = 339.014934(4)$^{\circ}$, Dec = 33.975872(3)$^{\circ}$. We apply shifts of dRA = 0.299'' and dDec = $-0.440$'' with \textsc{wcsupdate}. The drawback of having only one point with which to do the correction is that we can not correct for potential effects of rotation.
After applying the bore-sight correction the position of the core of NGC 7319 differs from the 2MASS position by 0.03''. This is consistent with the uncertainty in the determination of the X-ray source on the CCD by \textsc{wavdetect}. We adopt this as the uncertainty on the bore-sight corrected \textit{Chandra} positions.
The corrected position for the ULX is RA = 339.02917(2)$^{\circ}$, Dec = 33.97476(2)$^{\circ}$ in the 2MASS frame. These uncertainties stem only from determining the positions of the ULX on the CCD and tying the \textit{Chandra} frame to the 2MASS frame (and thus do not take into account any uncertainty associated with tying the 2MASS sources to the International Celestial Reference System [ICRS]). All quoted uncertainties correspond to 1-$\sigma$ errors on single parameters.

We extract all counts in a circle with a radius of 6 pixels around the source using the \textsc{CIAO}-task \textsc{specextract}. The 6-pixel radius contains approximately 95\% of the energy of the source. We use \textsc{arfcorr} to apply an aperture correction to the auxiliary response file (arf) to give us the correct source flux. To calculate the background we also extract the counts in a circle with a radius of 80 pixels away from any sources but on the same CCD chip. The observed background-subtracted number of photons in the 0.3--7.0 keV range is 162. The predicted number of background photons in the source extraction region is 4. We use \textsc{Xspec} version 12.6.0q to fit an absorbed power law to the counts per second per keV using C-statistics \citep{cash79} modified to account for the subtraction of background counts, the so called W-statistics\footnote{see http://heasarc.gsfc.nasa.gov/docs/xanadu/xspec/manual/}.
The best-fit values are listed in Table \ref{tab1}. These values give a 0.5-10.0 keV luminosity of $2 \times 10^{40}$ erg s$^{-1}$ at the distance of the galaxy of 90.75 Mpc \citep{crook07}.

\begin{table*}
\begin{minipage}{15cm}
\caption{The best-fit parameters for an absorbed power law fitted to the spectrum of NGC 7319-X4 and for an absorbed power law with pileup fitted to the spectrum of NGC 5474-X1.}
 \begin{tabular}{l c c}
 \hline
parameter & NGC 7319-X4 & NGC 5474-X1 \\
\hline
photon index & 1.8 $\pm$ 0.3 & 1.59 $\pm$ 0.06  \\
normalization (0.5-10.0 keV) & $(2.0 \pm 0.2) \times 10^{-14}$ erg cm$^{-2}$ s$^{-1}$ & $(3.4 \pm 0.1) \times 10^{-12}$ erg cm$^{-2}$ s$^{-1}$\\
column density & $(1.7 \pm 0.8) \times 10^{21} \textrm{cm}^{-2}$ & $(9.7 \pm 1.5) \times 10^{20}$ cm$^{-2}$\\
grade morphing parameter $\alpha$ & -- & 0.64 $\pm 0.02$ \\
 \hline
 \end{tabular}
\label{tab1}
\end{minipage}
\end{table*}

%\begin{table}
%\caption{The best-fit parameters for an absorbed power law fitted to the spectrum of NGC 7319-X4.}
% \begin{tabular}{l c}
% \hline
%parameter & best-fit value \\
%\hline
%photon index &  1.8 $\pm$ 0.3\\
%normalization (0.5-10.0 keV) & $(2.0 \pm 0.2) \times 10^{-14}$ erg cm$^{-2}$ s$^{-1}$\\
%column density & $(1.7 \pm 0.8) \times 10^{21} \textrm{cm}^{-2}$\\
% \hline
% \end{tabular}
%\label{tab4}
%\end{table}

We perform the same analysis on the shorter observation (with obsid 789). The observed background-subtracted number of photons is 20. We fit the same model to this dataset and find that the spectrum is consistent with the same model and flux.

To derive the visual extinction from the hydrogen column density we use the relation of \citet{predehl95} between $N_H$ and optical extinction. We find a visual extinction $A_V = 0.9 \pm 0.4$ magnitudes. This value is higher than the Galactic foreground extinction in the direction of NGC 7319 provided by NED\footnote{Nasa/ipac Extragalactic Database, http://ned.ipac.caltech.edu/}, $A_V = 0.26$ mag \citep{schlegel98}. However, the dust maps of \citet{schlegel98} have the same resolution as for IRAS, leaving room for local fluctuations. 

\subsubsection{NGC 5474}
\begin{figure*}
 \hbox{
\includegraphics[width=0.4\textwidth]{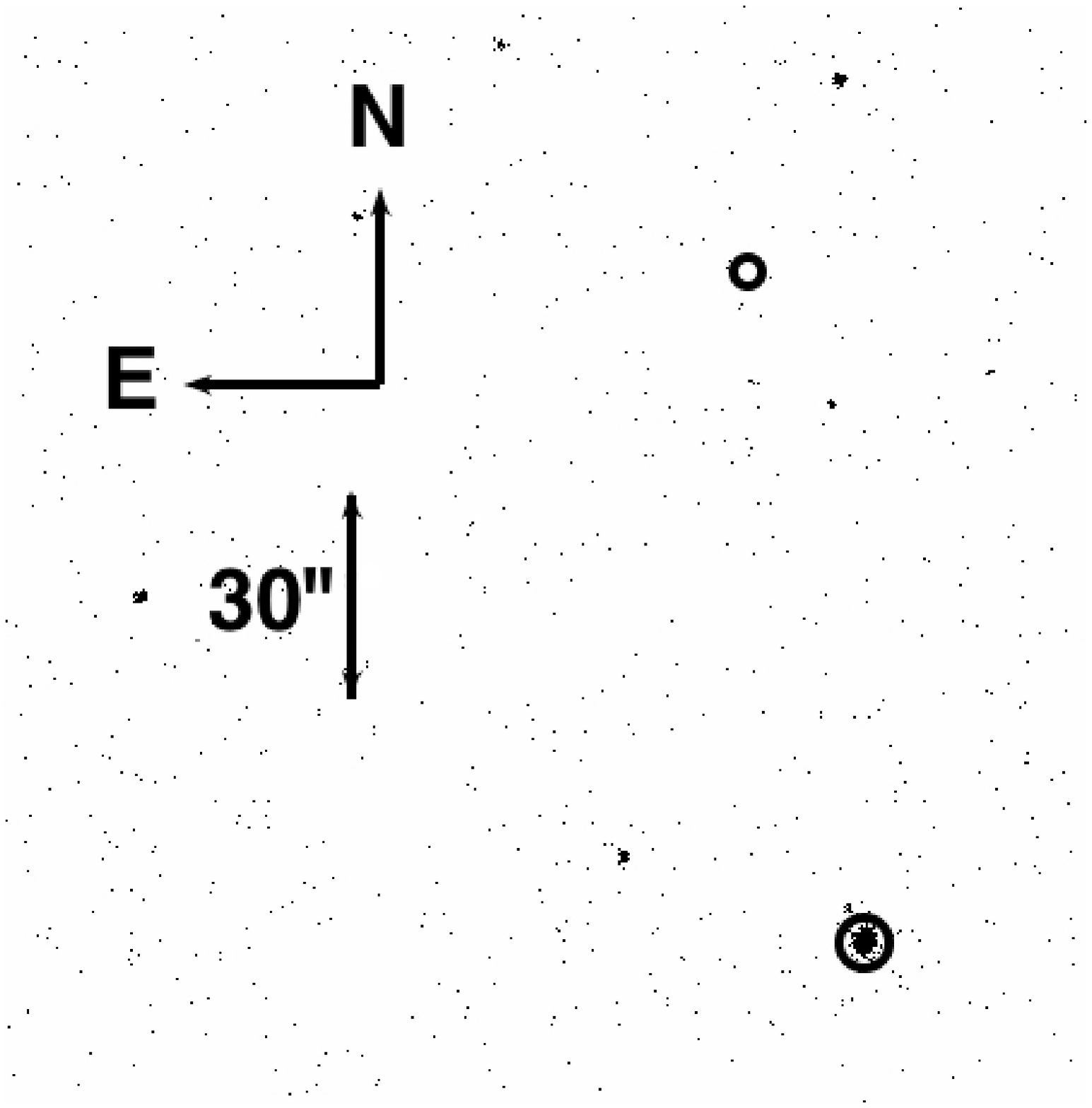}
\includegraphics[width=0.6\textwidth]{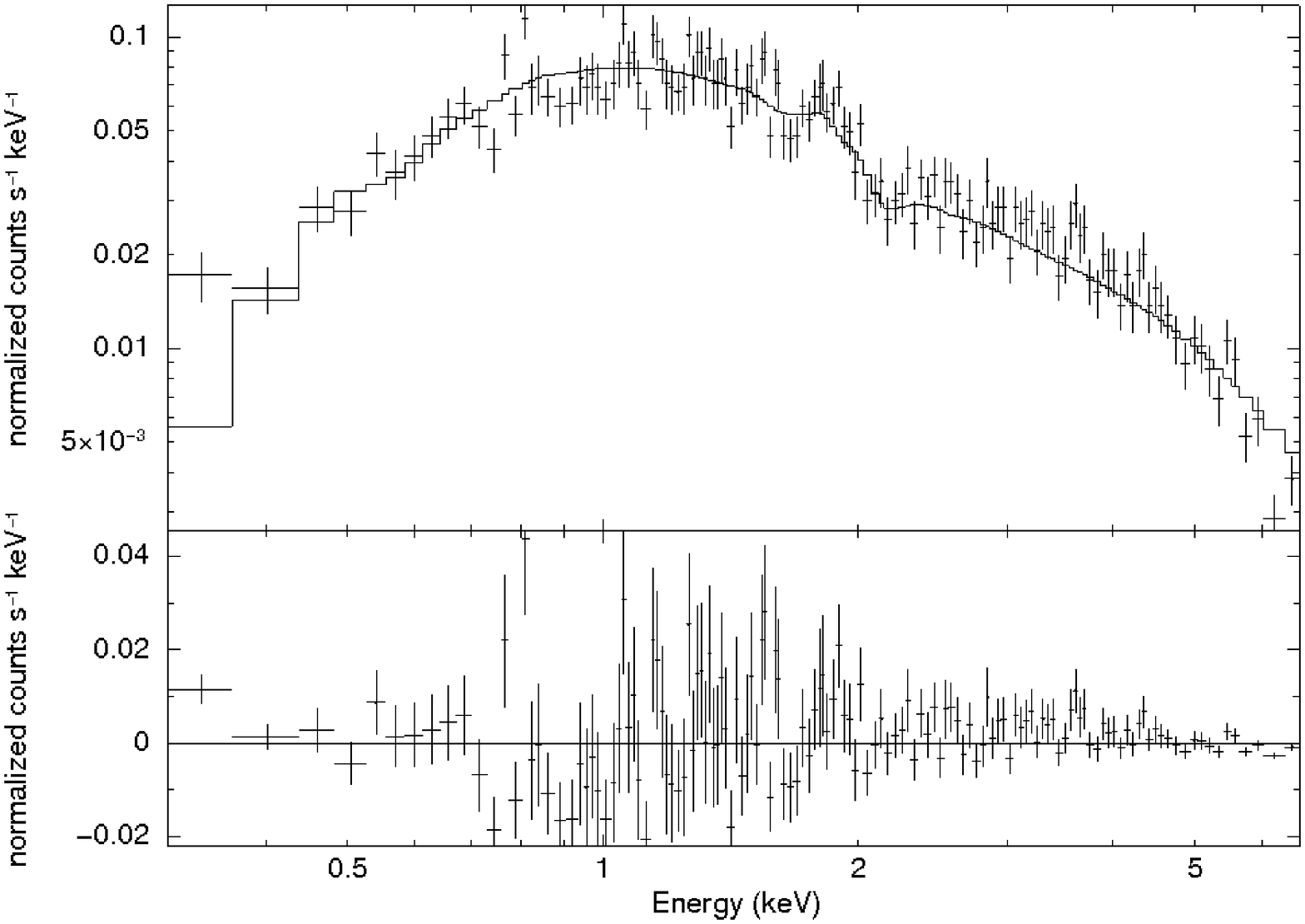}
}
\caption{\emph{Left panel:} The \textit{Chandra} image with the ULX (lower circle) and position of the core of NGC 5474 (upper circle). \emph{Right panel:} The X-ray spectrum of NGC 5474-X1 with fitted model and residuals. The counts are extracted from the observation with ID 9546 (see Table \ref{tab2}).}\label{chan5474}
\end{figure*}

We follow the same procedure for NGC 5474. For the bore-sight correction we use the star USNO B1.0 1436-0232144. Its position is RA = 211.26471(3)$^{\circ}$, Dec = 53.63914(3)$^{\circ}$. It has no measured proper motion. The position of the corresponding X-ray source is RA = 211.26499(2)$^{\circ}$, Dec = 53.63916(2)$^{\circ}$. After correcting the astrometry the offset between the reference star and its USNO position is 0.15''. We adopt this as the uncertainty of the X-ray positions with respect to the USNO B1.0 frame. The corrected position of the ULX is RA = 211.24859(3)$^{\circ}$, Dec = 53.63584(3)$^{\circ}$ in the USNO B1.0 frame.

The net number of counts from the source is 5219 in the 0.3--7.0 keV range, with a count rate of $0.175 \pm 0.002$ counts per second. The background level is less than 0.05 \%, corresponding to less than 2 photons. We use the \textsc{Ftools} task \textsc{grppha} to rebin the spectrum with at least 30 counts in each bin. We fit an absorbed power law (\textsc{wabs} $\times$ \textsc{pegpwrlw}) to the data using $\chi^2$ statistics. Because of the high count rate we also add the pileup model \citep{davis01} to the fit function. The best-fit parameters are listed in Table \ref{tab1} and the X-ray spectrum with the fitted model is plotted in Figure \ref{chan5474}. The ULX has an 0.5--10.0 keV X-ray luminosity of $1.9 \times 10^{40}$ erg s$^{-1}$ at the distance of the galaxy of 6.8 Mpc \citep{drozdovsky00}. 

We use the same extraction regions on the shorter exposure with ID 7086. The background-subtracted number of source counts is 182, too few to use $\chi^2$ statistics. We use W-statistics instead. The data are consistent with the same model as the longer exposure, but the flux is significantly lower: in the shorter exposure the 0.5--10 keV flux is $(0.7 \pm 0.1) \times 10^{-12}$ erg cm$^{-2}$ s$^{-1}$, almost a  factor 5 lower than in the later, longer exposure. The 0.5--10.0 keV X-ray luminosity is $3.9 \times 10^{39}$ erg s$^{-1}$

%\begin{table}
%\caption{The best-fit parameters for an absorbed power law with pileup fitted to the spectrum of NGC 5474-X1. The reduced $\chi^2$  = 1.56 for 136 degrees of freedom.}
% \begin{tabular}{l c c}
% \hline
%parameter & NGC 7319-X4 & NGC 5474-X1 \\
%\hline
%photon index & 1.8 $\pm$ 0.3 & 1.59 $\pm$ 0.06  \\
%normalization (0.5-10.0 keV) & $(2.0 \pm 0.2) \times 10^{-14}$ erg cm$^{-2}$ s$^{-1}$ & $(3.4 \pm 0.1) \times 10^{-12}$ erg cm$^{-2}$ s$^{-1}$\\
%column density & $(1.7 \pm 0.8) \times 10^{21} \textrm{cm}^{-2}$ & $(9.7 \pm 1.5) \times 10^{20}$ cm$^{-2}$\\
%grade morphing parameter $\alpha$ & N.A. & 0.64 $\pm 0.02$ \\
% \hline
% \end{tabular}
%\label{tab3}
%\end{table}

Again using \citet{predehl95} we find an absorption $A_V = 0.54 \pm 0.08$ magnitudes. This is significantly more than the Galactic absorption in the direction of NGC 5474 of $A_V = 0.036$ mag \citep{schlegel98}.

\subsection{HST observations}

\subsubsection{NGC 7319}
We analyse archival HST data of NGC 7319 obtained with the Wide-Field and Planetary Camera 2 (WFPC2) on 1999-06-17 as part of the proposal GO-6596. The observations consist of 1 $\times$ 1600 s and 3 $\times$ 1700 s exposures with the F450W (wide B-band) filter, 4 $\times$ 800 s exposures with the F569W (WFPC2 V-band) filter and 4 $\times$ 500 s exposures with the F814W (WFPC2 I-band) filter. The position of the ULX is situated on WFC chip 2.

\begin{figure*}
\hbox{
\includegraphics[width=0.5\textwidth]{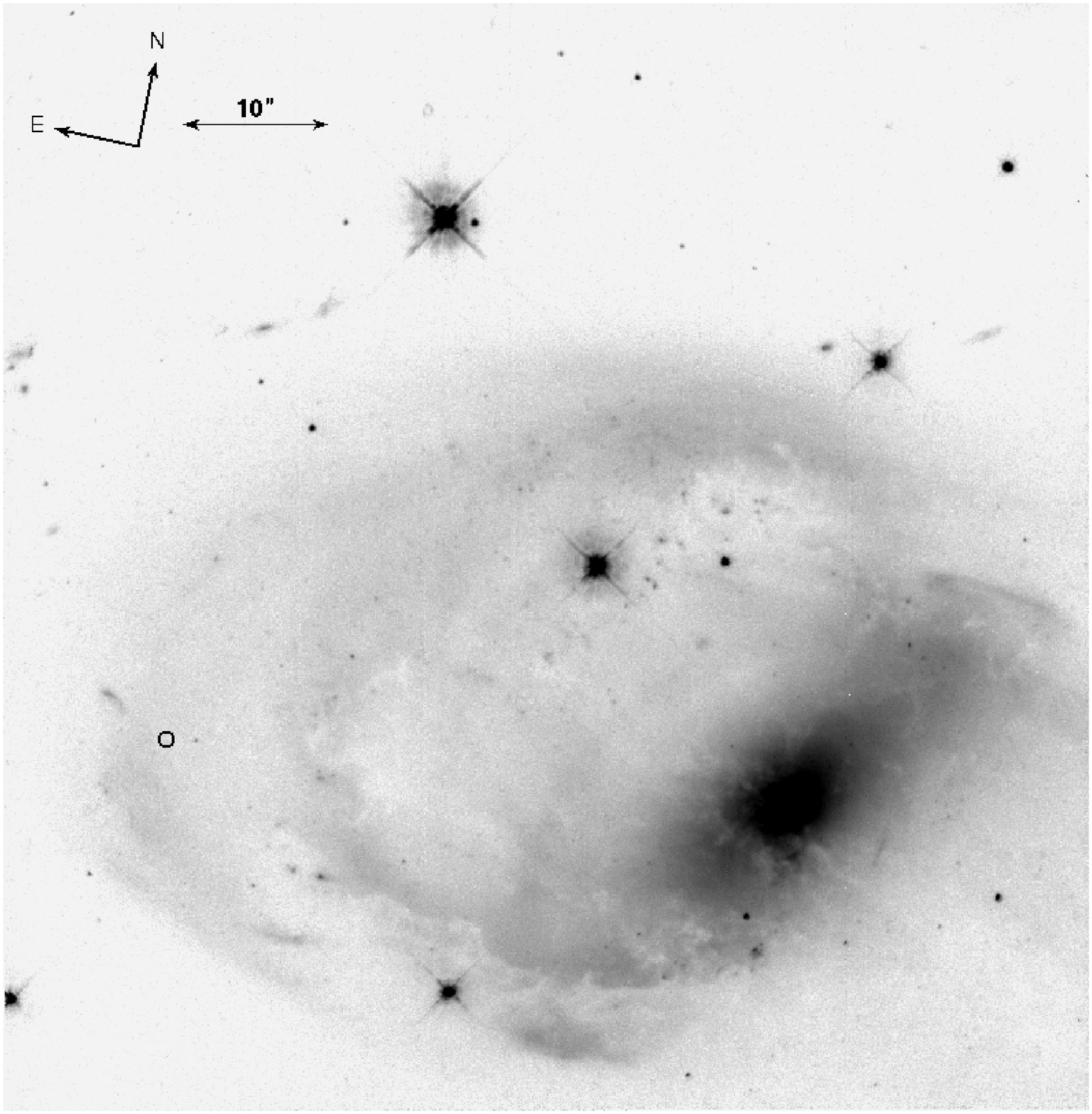}
 \includegraphics[width=0.5\textwidth]{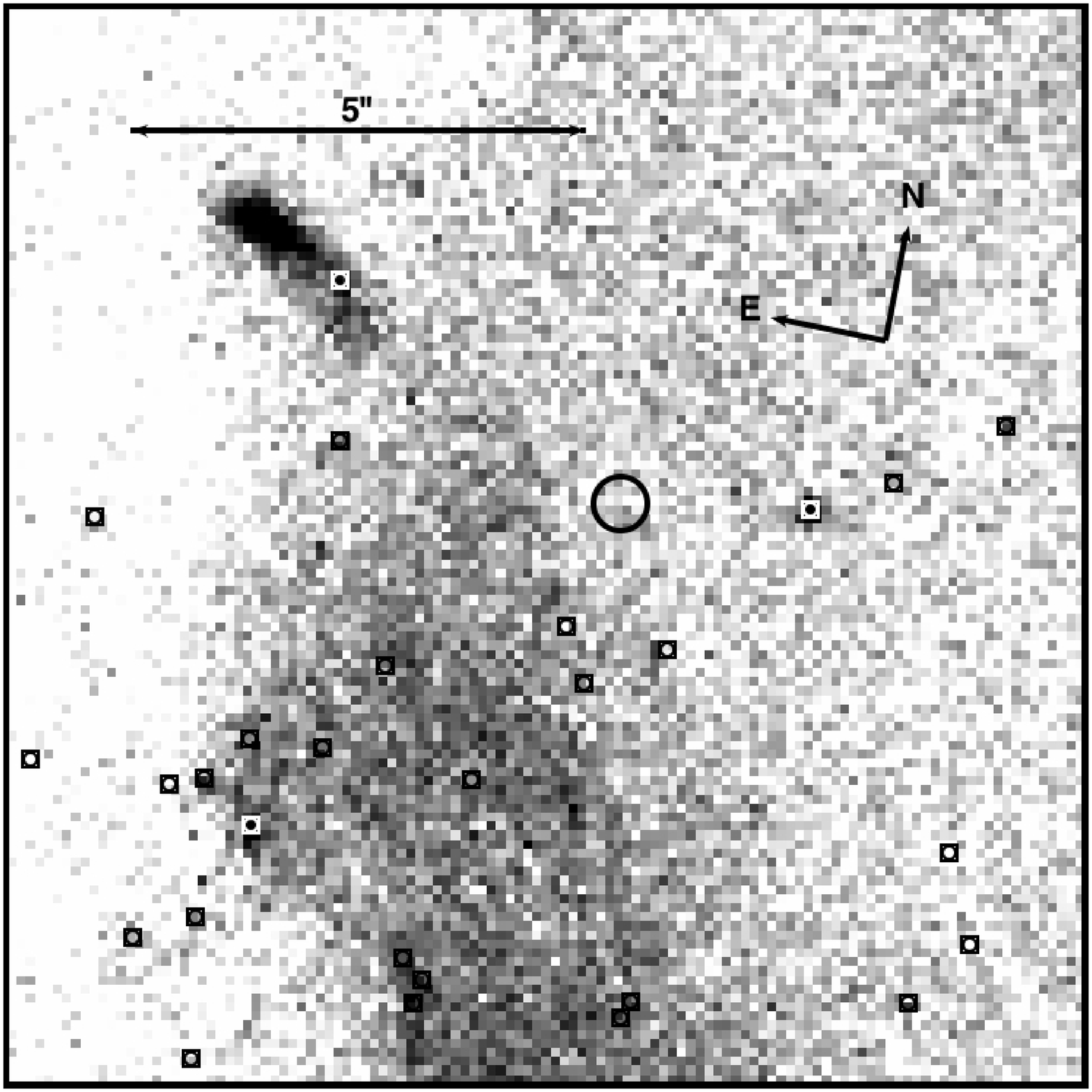}
 }
\caption{\emph{Left panel:} The 3200 s combined WFPC2 F569W image of NGC 7319 with a 0.5'' circle around the position of the X-ray source. \emph{Right panel:} Zoom-in of the stacked WFPC2 F569W images with the 0.3'' 1-$\sigma$ error circle around the position of the X-ray source and positions of stars from hstphot (boxcircles). Some of the boxcircles seem not to contain a star, these are stars that were only detected in the B- or I-band images.}
\label{hst7319}
\end{figure*}

We use the WFPC2 photometry package \textsc{Hstphot} \citep{dolphin00} to find stars in the HST data. Following the \textsc{Hstphot} User's Guide\footnote{http://purcell.as.arizona.edu/hstphot/} we first mask bad pixels and cosmics in the data image with the \textsc{mask}-task that uses the data quality image. To produce a single deep image in the three different bands we align the images using the \textsc{Iraf} task \textsc{imalign} and combine the aligned images with the \textsc{Hstphot} task \textsc{coadd}. We use the Sloan Digital Sky Survey (SDSS) positions of five unsaturated stars on the WFC2 chip to improve the astrometry with the \textsc{GAIA} software tool. The systematic uncertainty of the SDSS positions is $\sim0.25''$ \citep{pier03} and the uncertainty in the fit is $0.07''$, resulting in a total uncertainty of $\sim0.25''$. The total uncertainty of the X-ray position on the HST image also includes the systematic uncertainty of 2MASS (0.2'', \citealt{skrutskie06}), the statistical X-ray bore-sight uncertainty (0.03'') and the uncertainty in localizing the \textit{Chandra} source on the chip (0.1''). Adding these values in quadrature yields a 1-$\sigma$ uncertainty  of 0.3''.

We then use \textsc{Hstphot}'s main task -- also called \textsc{hstphot} -- to get photometric information of all the stars in the images. It converts the flight system magnitudes to WFPC2 standard magnitudes using the transformations described in \citet{holtzman95}. We use the `local sky determination' option of \textsc{hstphot} because the background varies across the chip due to the presence of a variable amount of unresolved stars in NGC 7319. The program returns several stars around the \textit{Chandra} position of the X-ray source, but none closer than 0.8'' (see Figure \ref{hst7319}). The probability that this closest source is the optical counterpart to the ULX is negligible.

We also use \textsc{Hstphot}'s function to add artificial stars to an image, with the B-band magnitude ranging from 24 to 30 and the B-I color from $-1$ to 6. The U-, V- and R-band magnitudes are also estimated. Using this method we estimate the limiting magnitude of the images near the position of the ULX as the magnitude of the faintest stars that are detected at the 5-$\sigma$ level. We find upper limits of WFPC2 B-band magnitude $< 28$, WFPC2 V-band magnitude $< 27$ and WFPC2 I-band magnitude $< 26$.

\subsubsection{NGC 5474}
For NGC 5474 we analyse archival HST/WFPC2 data obtained on 2009-02-09 as part of proposal GO-11966 in two filters: 3 $\times$ 600 s with F656N (narrow-band H$\alpha$) and 4 $\times$ 1100 s with F336W (WFPC2 U-band). Due to the shorter exposure and the narrow band of the H$\alpha$ filter, the observations are not sensitive enough to detect sufficient sources to tie them to the U-band observations. Therefore, we use only the broad U-band observations, datasets ub9u3004m, ub9u3005m, ub9u3006m and ub9u3007m. The ULX position is situated on WFC chip 3. 

\begin{figure*}
\hbox{
\includegraphics[width=0.5\textwidth]{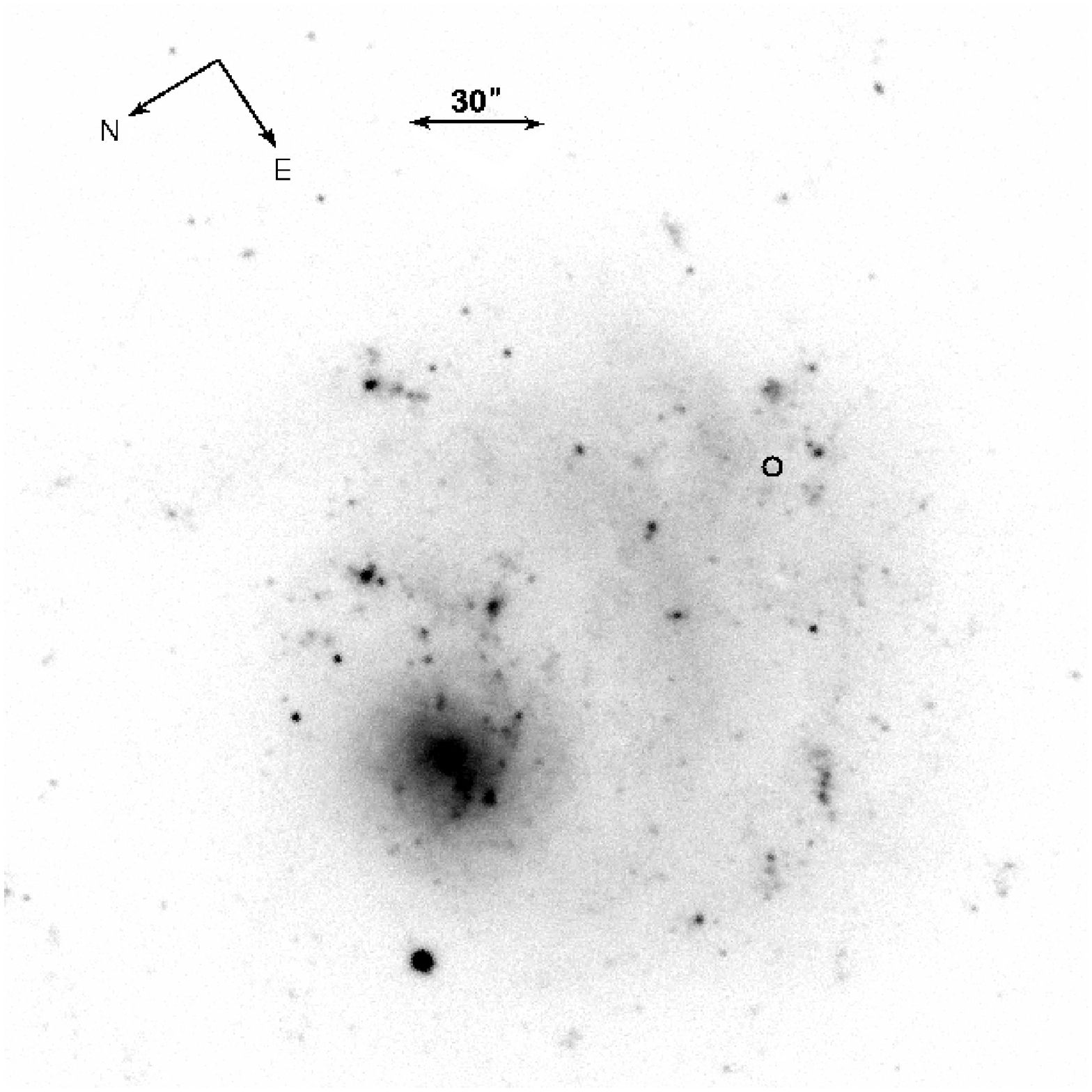}
\includegraphics[width=0.5\textwidth]{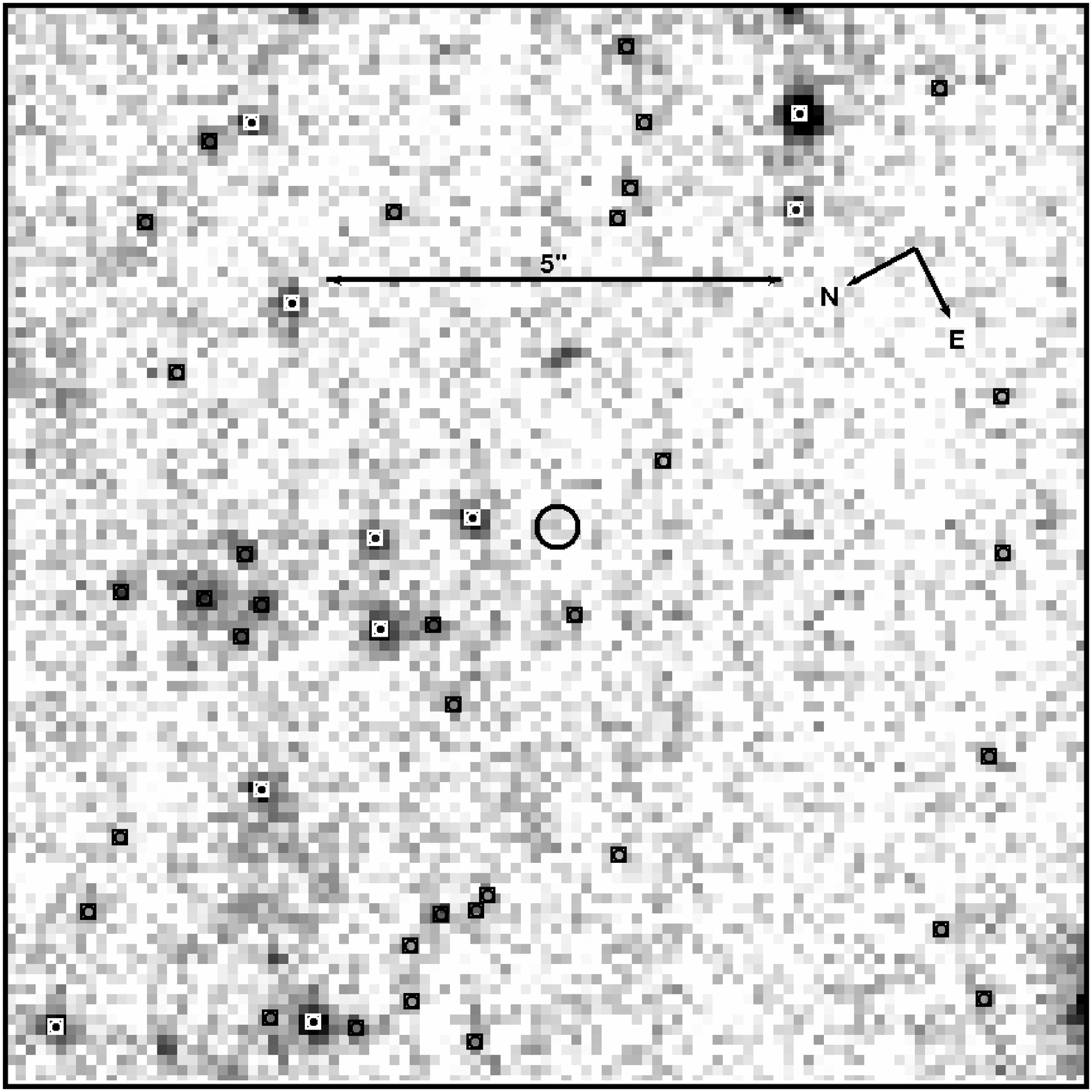}
}
\caption{\emph{Left panel:} The SDSS g-band image of NGC 5474 with a 2'' circle around the position of the X-ray source. \emph{Right panel:} Zoom-in of the stacked WFPC2 F336W image with the 0.2'' 1-$\sigma$ error circle around the position of the X-ray source and positions of stars from hstphot (boxcircles).}
\label{hst5474}
\end{figure*}

We process the data in the same way as for NGC 7319. To improve the astrometry of HST we use four stars on the WFC3 chip that are also in the USNO B1.0 catalog. The root-mean-square value of the fit is 0.17''. Because we are only interested in the relative positions of sources in the \textit{Chandra} and HST images and not in the absolute astrometry we do not have to take the systematic uncertainty of the USNO B1.0 positions with respect to the ICRS into account. Adding the uncertainty on the \textit{Chandra} position, the total uncertainty of the position of the ULX on the HST image becomes $\sqrt{0.17^2 + 0.15^2} \approx 0.22''$. The nearest source in HST that \textsc{hstphot} finds lies at a distance of 0.8'' (see Figure \ref{hst5474}), this is more than 3.5 $\sigma$ away from the X-ray position.

We use the simulated stars option of \textsc{hstphot} to investigate the limiting magnitude for this observation near the ULX position. The faintest star we detect at more than 5 $\sigma$ has an apparent WFPC2 U-band magnitude of 23.7. We take this as the limiting magnitude for these U-band observations.

\subsection{Limiting absolute magnitudes}
We do not find optical counterparts to the ULXs in NGC 7319 and NGC 5474. We can, however, calculate limiting absolute magnitudes for the possible companions. 
The distance modulus to NGC 7319 is 34.8. Using values from \citet{cardelli89} to transform $A_V$ to the extinction in the filters that we used we find that the extinction due to absorption is $1.1 \pm 0.5$ magnitudes in the F450W image, $0.9 \pm 0.4$ magnitudes in the F569W image and $0.5 \pm 0.2$ in the F814W image. The limiting absolute magnitude in the WFPC2 B-band is $-7.9$. In the WFPC2 V-band it is $-8.7$ and in the WFPC2 I-band $-9.3$.

The extinction in the U-band for NGC 5474 is $A_U = 0.9 \pm 0.2$ magnitudes. The distance modulus to NGC 5474 is 29.15 \citep{drozdovsky00}. The limiting absolute magnitude in the WFPC2 U-band is then $-6.4$. 

There are uncertainties present in the extinction due to absorption, the distance modulus and the limiting magnitudes of the observations. Of those we only know the uncertainty in the extinction. As an indication for the uncertainties in the distance modulus we use the standard deviations given by NED. The distance to NGC 7319 is based only on redshift. The uncertainty in the distance modulus due to uncertainties in the redshift measurement is 0.15, and there is also an (unknown) systematic uncertainty involved in deriving the distance from the redshift caused by e.g. potential proper velocities with respect to the Hubble flow. For NGC 5474 NED gives a standard deviation of the distance modulus of 0.23 based on five individual distance measurements. We conclude that the uncertainties in the distance modulus and extinction are equally important and the total uncertainty on the limiting magnitudes is of the order of 0.5 magnitude.

\section{Discussion}
Using bore-sight corrected \textit{Chandra} images we determine accurate positions for the ULXs NGC 7319-X4 and NGC 5474-X1. We also improve the astrometry of HST observations of these galaxies and search for optical sources at the ULX positions. The position of NGC 7319-X4 is RA = 339.02917$^{\circ}$, Dec = 33.97476$^{\circ}$, with a 1-$\sigma$ error circle of 0.3'' on the HST image. We find the nearest star at 0.8'' from this position -- at the distance of NGC 7319 (90.75 Mpc) this corresponds to $\sim$ 350 pc.
The position of NGC 5474-X1 is RA = 211.24859$^{\circ}$, Dec = 53.63584$^{\circ}$, with a 1-$\sigma$ error circle on the HST image of 0.22''. Again the nearest star that we find lies at 0.8'' from this position. This corresponds to $\sim$ 25 pc at the distance of NGC 5474.
We do not find the optical counterparts to the ULXs in NGC 5474 and NGC 7319 in the HST data, but using our limiting absolute magnitudes we can put constraints on their counterparts. To constrain the spectral type of a possible companion star we use table 15.7 from Allen's astrophysical quantities \citep{cox_book}. This table contains absolute V-magnitude and color calibrations for the MK spectral classes. We use the values from the WFPC2 Cookbook\footnote{www.stsci.edu/hst/wfpc2/analysis/wfpc2\_cookbook.html} to convert these UBVRI magnitudes to WFPC2 standard magnitudes.

\subsubsection{NGC 7319}
The limiting magnitudes for the source in NGC 7319 are $-7.9$ in the WFPC2 B-band, $-8.7$ in the WFPC2 V-band and $-9.3$ in the WFPC2 I-band. The brightest stars in the WFPC2 B- and V-band are O9 supergiants with a B-band magnitude of $-7.5$ and a V-band magnitude of $-6.5$, and in the WFPC2 I-band the brightest stars are M5 supergiants with a magnitude of $-8.7$. None of these would have been detected in the HST images. With our limiting absolute magnitudes we can not exclude any type of single star as counterpart.

If the ULX is an accreting compact object it will probably have an accretion disc. These can also be visible in optical light -- in fact, disc emission may dominate the emission in the blue part of the spectrum. Can our limiting absolute magnitudes constrain the size of such a disc? \citet{vanparadijs94} found a relation between $\Sigma$ and the absolute V-band magnitude of low-mass X-ray binaries (LMXBs). $\Sigma$ depends on the Eddington ratio and period of the system as $\Sigma = (L_X/L_{Edd})^{1/2} (P/1 hr)^{2/3}$.  Using this relation and the limiting V-band magnitude of $-8.7$ yields an upper limit for $\log(\Sigma)$ of 4.5 (with the caveat that we are extrapolating this relation from $\log(\Sigma) = 2.01$, the maximum value in the sample of \citet{vanparadijs94}). This is not very constraining -- if we assume that the system is accreting at the Eddington limit we find a maximum period of 640 years, much longer than that of any LMXB. This does not give us a useful constraint on the size of a possible accretion disc.

At these high X-ray luminosities X-ray irradiation of the donor star and accretion disc is likely to play an important role (\citealt{copperwheat05}, \citealt{copperwheat07}, \citealt{patruno08}). \citet{copperwheat05} give absolute V-magnitudes for an irradiated O5V star and G0I star with irradiated disc around a 10, 100 and 1000 M$_{\odot}$ black hole (their figure 6). The ULX in NGC 7319 has a hardness ratio of $\sim$ 1.4. An absolute limiting V-band magnitude of $-8.7$ means that an O5V star would not have been detected regardless of the mass of the black hole, but a G0 supergiant around a 100 or 1000 M$_{\odot}$ black hole would have been visible. 

The most X-ray bright supernovae, type IIn, are also bright at optical wavelengths. The average intrinsic B-band magnitude at maximum brightness of these SN is $M_B = -19.15 \pm 0.92$ \citep{richardson02}, and they fade on timescales of years \citep{filippenko97}. Figure 4 of \citet{tsvetkov08} compares the lightcurves of 5 SN IIn, showing that M$_V \leq -15$ for more than 400 days. The X-ray luminosity peaks around 400-1000 day after the explosion. 
This makes the supernova scenario unlikely for this source. The HST observations were made a year before the first X-ray observation. If the supernova had gone off before the HST observation, it would have been detected in the B-band images. So the only possibility is that it went off after the HST observations, but in time for it to be bright in X-rays a year later, leaving a very short period of time in which the explosion could have occurred. 

Since NGC 7319 is a Seyfert 2 galaxy and hence has a detected nuclear X-ray source, the ULX is probably not a recoiling supermassive black hole.

Another scenario that we have to investigate is that the X-ray source may be a background quasar. If the source is situated behind NGC 7319, we expect a significant contribution of that galaxy to the column density of neutral hydrogen on top of the Galactic value. We assume a typical disc height of 100 parsec and an average density of 1 particle per cm$^3$ \citep{ferriere01}. The contribution of a face-on spiral galaxy to the column density is then $\sim 3 \times 10^{20}$ cm$^{-2}$, which gives a visual extinction $A_V \approx 0.2$ (using the relation of \citealt{predehl95}). The foreground Galactic extinction in the direction of NGC 7319 is $A_V = 0.26$; adding the extra extinction from NGC 7319 gives $A_V \approx 0.46$. This is lower than, although consistent with, the value that we find from the X-ray spectrum, $A_V = 0.9 \pm 0.4$.
Variability in the X-ray spectrum can also help to exclude the AGN scenario. We do not find significant variability between the two \textit{Chandra} observations; therefore we cannot refute the possibility that this source is a background AGN based on X-ray variability.
Another clue about the background nature of the source can be gained from the X-ray to optical flux ratio. Most \textit{Chandra}-selected AGN have an X-ray (0.5-2.0 keV) to optical (R-band) flux ratio smaller than 10 (\citealt{barger03}, \citealt{laird09}). For this source however, this ratio is larger than 70 in the V-band, larger than 74 in the B-band and larger than 90 in the I-band. The optical flux limits have been corrected for absorption. These values make the background AGN scenario less likely for this source.

\subsubsection{NGC 5474}
For NGC 5474 we have a limiting absolute magnitude of $-6.4$ in the WFPC2 U-band. This means that the counterpart can not be an O type main sequence star, nor an A-type or earlier supergiant. Those stars have a WFPC2 U-band magnitude lower than $-6.4$ so they should have been detected in the HST image.

Assuming that $F_{\lambda} \propto \lambda^{-2}$ for accretion discs gives M$_U$ - M$_V \approx -1$, yielding a limiting V-band magnitude of approximately $-5.4$ for this source. Using the relation of \citet{vanparadijs94}, this limiting magnitude infers a maximum value of $\log\Sigma$ of 3.1, or, assuming that the ULX is accreting at its Eddington limit, a maximum period of $\sim$ 5 years (again, we are extrapolating the relation between $\Sigma$ and M$_V$). This is such a high upper limit that an LMXB accretion disc would not be detected in these images. X-ray irradiation of the disc and donor star may increase the optical luminosity of the system.

The combination of an increasing X-ray luminosity and the non-detection of an optical counterpart in the U-band three years after the first \textit{Chandra} observation makes it very unlikely that NGC 5474-X4 is a supernova. To make that scenario work the first \textit{Chandra} observation (in September 2006) would have had to catch the very steep rise of the X-ray luminosity about one year after the supernova \citep{immler03}, but in that case it would have been visible in the U-band HST image three years later \citet{tsvetkov08}. On top of that, a SN IIn in NGC 5474 would have a peak apparent magnitude m$_V \approx 10$. It seems very unlikely that such a source went unnoticed.

Since we do not observe a nuclear X-ray source in NGC 5474 we cannot exclude the recoiling SMBH scenario on that ground. In the case of a recoiling SMBH the optical counterpart would probably be related to the material that remains gravitationally bound to the black hole -- the accretion disc, broad line region and part of the nuclear star cluster -- and its luminosity would be comparable to that of an AGN. The absolute magnitudes of AGN vary, but the faintest that have been reported thus far have absolute magnitudes M$_B$ = -9.8 for NGC 4395 \citep{filippenko89} and -11.6 for NGC 3031/M81 \citep{ho97}. With a limiting U-band magnitude of $-6.4$ NGC 5474-X1 is much fainter than these faintest AGNs, making it unlikely that it is a recoiling SMBH.

The Galactic absorption in the direction of NGC 5474 is only $A_V = 0.036$ mag, while we find an absorption of $A_V = 0.54 \pm 0.08$ mag from the X-ray spectrum. This means there is significant absorption on top of the Galactic foreground, possibly local to the source or because the source is situated behind NGC 5474. The two \textit{Chandra} observations, spaced 15 months apart, show similar spectra but with a factor 5 increase in luminosity towards the later observation. This does not exclude a background AGN. It does mean that this is indeed a single source instead of e.g. a group of LMXBs. 
However, the X-ray (0.5-2.0 keV) to optical (U-band) flux ratio of this source is larger than 500, while typically AGN have a ratio smaller than 10 in the R-band. This argues strongly against the background AGN scenario.

\subsection{Conclusions}
We can not exclude any type of single star as the companion of the ULX in NGC 7319. The limiting magnitudes for this source also allow an LMXB with accretion disc. \citet{copperwheat05} calculated V-band magnitudes for X-ray irradiated discs and donor stars. By comparing these with our limiting magnitude we can exclude a G0 supergiant around an IMBH of more than 100 M$_{\odot}$. The source could also be a supernova, although this is only possible if it went off in a short period of time right after the HST observations. The recoiling black hole scenario is not very likely for this source, given that NGC 7319 also has an AGN. The source could be a background AGN but the X-ray to optical flux ratios argue against this scenario.

The companion of the ULX in NGC 5474 cannot be a blue supergiant or O-type main sequence star. Other stellar types and an LMXB with accretion disc are still an option. The source is probably not a supernova, since the increasing X-ray luminosity and non-detection in the HST observation make this very unlikely. It can be a recoiling black hole, but then its optical luminosity is much lower than is usual for AGN. There is significant absorption on top of the Galactic foreground absorption in the direction of NGC 5474 which could point to a background object. However, the X-ray to optical flux ratio is much higher than usual for AGN, which argues strongly against this scenario. Finding the optical counterpart or measuring variability of the X-ray source can help decide whether or not this is a background object.

\section*{Acknowledgements}
MH and PGJ thank Richard Mushotzky for bringing NGC 7319-X4 to our attention. 
This research has made use of data obtained from the \textit{Chandra} Data Archive and the \textit{Chandra} Source Catalog, and software provided by the \textit{Chandra} X-ray Center (CXC) in the application package CIAO. This paper is based on observations made with the NASA/ESA Hubble Space Telescope, obtained from the data archive at the Space Telescope Institute. STScI is operated by the association of Universities for Research in Astronomy, Inc. under the NASA contract NAS 5-26555.

 \bibliographystyle{mn2e}
 \bibliography{bibliography}

\begin{thebibliography}{}

\bibitem[\protect\citeauthoryear{{Baker}, {Boggs}, {Centrella}, {Kelly},
  {McWilliams}, {Miller} \& {van Meter}}{{Baker} et~al.}{2008}]{baker08}
{Baker} J.~G.,  {Boggs} W.~D.,  {Centrella} J.,  {Kelly} B.~J.,  {McWilliams}
  S.~T.,  {Miller} M.~C.,    {van Meter} J.~R.,  2008, \apjl, 682, L29

\bibitem[\protect\citeauthoryear{{Barger}, {Cowie}, {Capak}, {Alexander},
  {Bauer}, {Fernandez}, {Brandt}, {Garmire} \& {Hornschemeier}}{{Barger}
  et~al.}{2003}]{barger03}
{Barger} A.~J.,  {Cowie} L.~L.,  {Capak} P.,  {Alexander} D.~M.,  {Bauer}
  F.~E.,  {Fernandez} E.,  {Brandt} W.~N.,  {Garmire} G.~P.,    {Hornschemeier}
  A.~E.,  2003, \aj, 126, 632

\bibitem[\protect\citeauthoryear{{Belczynski}, {Bulik}, {Fryer}, {Ruiter},
  {Valsecchi}, {Vink} \& {Hurley}}{{Belczynski} et~al.}{2010}]{belczynski10}
{Belczynski} K.,  {Bulik} T.,  {Fryer} C.~L.,  {Ruiter} A.,  {Valsecchi} F.,
  {Vink} J.~S.,    {Hurley} J.~R.,  2010, \apj, 714, 1217

\bibitem[\protect\citeauthoryear{{Cardelli}, {Clayton} \& {Mathis}}{{Cardelli}
  et~al.}{1989}]{cardelli89}
{Cardelli} J.~A.,  {Clayton} G.~C.,    {Mathis} J.~S.,  1989, \apj, 345, 245

\bibitem[\protect\citeauthoryear{{Cash}}{{Cash}}{1979}]{cash79}
{Cash} W.,  1979, \apj, 228, 939

\bibitem[\protect\citeauthoryear{{Colbert} \& {Miller}}{{Colbert} \&
  {Miller}}{2005}]{colbert05}
{Colbert} E.~J.~M.,  {Miller} M.~C.,  2005, in {M.~Novello, S.~Perez
  Bergliaffa, \& R.~Ruffini} ed., The Tenth Marcel Grossmann Meeting. On recent
  developments in theoretical and experimental general relativity, gravitation
  and relativistic field theories {Observational Evidence for Intermediate-Mass
  Black Holes in Ultra-Luminous X-Ray Sources}.
pp 530--+

\bibitem[\protect\citeauthoryear{{Copperwheat}, {Cropper}, {Soria} \&
  {Wu}}{{Copperwheat} et~al.}{2005}]{copperwheat05}
{Copperwheat} C.,  {Cropper} M.,  {Soria} R.,    {Wu} K.,  2005, \mnras, 362,
  79

\bibitem[\protect\citeauthoryear{{Copperwheat}, {Cropper}, {Soria} \&
  {Wu}}{{Copperwheat} et~al.}{2007}]{copperwheat07}
{Copperwheat} C.,  {Cropper} M.,  {Soria} R.,    {Wu} K.,  2007, \mnras, 376,
  1407

\bibitem[\protect\citeauthoryear{{Cox}}{{Cox}}{2000}]{cox_book}
{Cox} A.~N.,  2000, {Allen's astrophysical quantities}

\bibitem[\protect\citeauthoryear{{Crook}, {Huchra}, {Martimbeau}, {Masters},
  {Jarrett} \& {Macri}}{{Crook} et~al.}{2007}]{crook07}
{Crook} A.~C.,  {Huchra} J.~P.,  {Martimbeau} N.,  {Masters} K.~L.,  {Jarrett}
  T.,    {Macri} L.~M.,  2007, \apj, 655, 790

\bibitem[\protect\citeauthoryear{Damour \& Gopakumar}{Damour \&
  Gopakumar}{2006}]{damour06}
Damour T.,  Gopakumar A.,  2006, Phys. Rev. D, 73, 124006

\bibitem[\protect\citeauthoryear{{Davis}}{{Davis}}{2001}]{davis01}
{Davis} J.~E.,  2001, \apj, 562, 575

\bibitem[\protect\citeauthoryear{{Dolphin}}{{Dolphin}}{2000}]{dolphin00}
{Dolphin} A.~E.,  2000, \pasp, 112, 1383

\bibitem[\protect\citeauthoryear{{Drozdovsky} \& {Karachentsev}}{{Drozdovsky}
  \& {Karachentsev}}{2000}]{drozdovsky00}
{Drozdovsky} I.~O.,  {Karachentsev} I.~D.,  2000, \aaps, 142, 425

\bibitem[\protect\citeauthoryear{{Ferrarese} \& {Merritt}}{{Ferrarese} \&
  {Merritt}}{2002}]{ferrarese02}
{Ferrarese} L.,  {Merritt} D.,  2002, ArXiv Astrophysics e-prints

\bibitem[\protect\citeauthoryear{{Ferri{\`e}re}}{{Ferri{\`e}re}}{2001}]{ferrie%
re01}
{Ferri{\`e}re} K.~M.,  2001, Reviews of Modern Physics, 73, 1031

\bibitem[\protect\citeauthoryear{{Filippenko}}{{Filippenko}}{1997}]{filippenko%
97}
{Filippenko} A.~V.,  1997, \araa, 35, 309

\bibitem[\protect\citeauthoryear{{Filippenko} \& {Sargent}}{{Filippenko} \&
  {Sargent}}{1989}]{filippenko89}
{Filippenko} A.~V.,  {Sargent} W.~L.~W.,  1989, \apjl, 342, L11

\bibitem[\protect\citeauthoryear{{Ho}, {Filippenko}, {Sargent} \& {Peng}}{{Ho}
  et~al.}{1997}]{ho97}
{Ho} L.~C.,  {Filippenko} A.~V.,  {Sargent} W.~L.~W.,    {Peng} C.~Y.,  1997,
  \apjs, 112, 391

\bibitem[\protect\citeauthoryear{{Holtzman}, {Burrows}, {Casertano}, {Hester},
  {Trauger}, {Watson} \& {Worthey}}{{Holtzman} et~al.}{1995}]{holtzman95}
{Holtzman} J.~A.,  {Burrows} C.~J.,  {Casertano} S.,  {Hester} J.~J.,
  {Trauger} J.~T.,  {Watson} A.~M.,    {Worthey} G.,  1995, \pasp, 107, 1065

\bibitem[\protect\citeauthoryear{{Immler} \& {Lewin}}{{Immler} \&
  {Lewin}}{2003}]{immler03}
{Immler} S.,  {Lewin} W.~H.~G.,  2003, in {K.~Weiler} ed., Supernovae and
  Gamma-Ray Bursters Vol.~598 of Lecture Notes in Physics, Berlin Springer
  Verlag, {X-Ray Supernovae}.
pp 91--111

\bibitem[\protect\citeauthoryear{{Jonker}, {Torres}, {Fabian}, {Heida},
  {Miniutti} \& {Pooley}}{{Jonker} et~al.}{2010}]{jonker10}
{Jonker} P.~G.,  {Torres} M.~A.~P.,  {Fabian} A.~C.,  {Heida} M.,  {Miniutti}
  G.,    {Pooley} D.,  2010, \mnras, 407, 645

\bibitem[\protect\citeauthoryear{{King}, {Davies}, {Ward}, {Fabbiano} \&
  {Elvis}}{{King} et~al.}{2001}]{king01}
{King} A.~R.,  {Davies} M.~B.,  {Ward} M.~J.,  {Fabbiano} G.,    {Elvis} M.,
  2001, \apjl, 552, L109

\bibitem[\protect\citeauthoryear{{Laird}, {Nandra}, {Georgakakis}, {Aird},
  {Barmby}, {Conselice}, {Coil}, {Davis}, {Faber}, {Fazio}, {Guhathakurta},
  {Koo}, {Sarajedini} \& {Willmer}}{{Laird} et~al.}{2009}]{laird09}
{Laird} E.~S.,  {Nandra} K.,  {Georgakakis} A.,  {Aird} J.~A.,  {Barmby} P.,
  {Conselice} C.~J.,  {Coil} A.~L.,  {Davis} M.,  {Faber} S.~M.,  {Fazio}
  G.~G.,  {Guhathakurta} P.,  {Koo} D.~C.,  {Sarajedini} V.,    {Willmer}
  C.~N.~A.,  2009, \apjs, 180, 102

\bibitem[\protect\citeauthoryear{{Liu}}{{Liu}}{2011}]{liu11}
{Liu} J.,  2011, \apjs, 192, 10

\bibitem[\protect\citeauthoryear{{Merritt}, {Schnittman} \&
  {Komossa}}{{Merritt} et~al.}{2009}]{merritt09}
{Merritt} D.,  {Schnittman} J.~D.,    {Komossa} S.,  2009, \apj, 699, 1690

\bibitem[\protect\citeauthoryear{{Motch}, {Pakull}, {Gris{\'e}} \&
  {Soria}}{{Motch} et~al.}{2011}]{motch11}
{Motch} C.,  {Pakull} M.~W.,  {Gris{\'e}} F.,    {Soria} R.,  2011,
  Astronomische Nachrichten, 332, 367

\bibitem[\protect\citeauthoryear{{Patruno} \& {Zampieri}}{{Patruno} \&
  {Zampieri}}{2008}]{patruno08}
{Patruno} A.,  {Zampieri} L.,  2008, \mnras, 386, 543

\bibitem[\protect\citeauthoryear{{Pier}, {Munn}, {Hindsley}, {Hennessy},
  {Kent}, {Lupton} \& {Ivezi{\'c}}}{{Pier} et~al.}{2003}]{pier03}
{Pier} J.~R.,  {Munn} J.~A.,  {Hindsley} R.~B.,  {Hennessy} G.~S.,  {Kent}
  S.~M.,  {Lupton} R.~H.,    {Ivezi{\'c}} {\v Z}.,  2003, \aj, 125, 1559

\bibitem[\protect\citeauthoryear{{Predehl} \& {Schmitt}}{{Predehl} \&
  {Schmitt}}{1995}]{predehl95}
{Predehl} P.,  {Schmitt} J.~H.~M.~M.,  1995, \aap, 293, 889

\bibitem[\protect\citeauthoryear{{Richardson}, {Branch}, {Casebeer}, {Millard},
  {Thomas} \& {Baron}}{{Richardson} et~al.}{2002}]{richardson02}
{Richardson} D.,  {Branch} D.,  {Casebeer} D.,  {Millard} J.,  {Thomas} R.~C.,
    {Baron} E.,  2002, \aj, 123, 745

\bibitem[\protect\citeauthoryear{{Schlegel}, {Finkbeiner} \&
  {Davis}}{{Schlegel} et~al.}{1998}]{schlegel98}
{Schlegel} D.~J.,  {Finkbeiner} D.~P.,    {Davis} M.,  1998, \apj, 500, 525

\bibitem[\protect\citeauthoryear{{Skrutskie}, {Cutri}, {Stiening}, {Weinberg},
  {Schneider}, {Carpenter}, {Beichman}, {Capps}, {Chester} et~al.,}{{Skrutskie}
  et~al.}{2006}]{skrutskie06}
{Skrutskie} M.~F.,  {Cutri} R.~M.,  {Stiening} R.,  {Weinberg} M.~D.,
  {Schneider} S.,  {Carpenter} J.~M.,  {Beichman} C.,  {Capps} R.,  {Chester}
  T.,    et~al., 2006, \aj, 131, 1163

\bibitem[\protect\citeauthoryear{{Swartz}, {Soria}, {Tennant} \&
  {Yukita}}{{Swartz} et~al.}{2011}]{swartz11}
{Swartz} D.~A.,  {Soria} R.,  {Tennant} A.~F.,    {Yukita} M.,  2011, \apj,
  741, 49

\bibitem[\protect\citeauthoryear{{Tsvetkov}}{{Tsvetkov}}{2008}]{tsvetkov08}
{Tsvetkov} D.~Y.,  2008, Peremennye Zvezdy, 28, 6

\bibitem[\protect\citeauthoryear{{van Paradijs} \& {McClintock}}{{van Paradijs}
  \& {McClintock}}{1994}]{vanparadijs94}
{van Paradijs} J.,  {McClintock} J.~E.,  1994, \aap, 290, 133

\bibitem[\protect\citeauthoryear{{Wiersema}, {Farrell}, {Webb}, {Servillat},
  {Maccarone}, {Barret} \& {Godet}}{{Wiersema} et~al.}{2010}]{wiersema10}
{Wiersema} K.,  {Farrell} S.~A.,  {Webb} N.~A.,  {Servillat} M.,  {Maccarone}
  T.~J.,  {Barret} D.,    {Godet} O.,  2010, \apjl, 721, L102

\bibitem[\protect\citeauthoryear{{Yang}, {Feng} \& {Kaaret}}{{Yang}
  et~al.}{2011}]{yang11}
{Yang} L.,  {Feng} H.,    {Kaaret} P.,  2011, \apj, 733, 118

\end{thebibliography}

\end{document}